# On digital simply connected spaces and manifolds: a digital simply connected 3-manifold is the digital 3-sphere

Alexander V. Evako

Npk Novotek, Laboratory of Digital Technologies. Moscow, Russia
e-mail: evakoa@mail.ru.

**Abstract**

In the framework of digital topology, we study structural and topological properties of digital n-dimensional manifolds. We introduce the notion of simple connectedness of a digital space and prove that if M and N are homotopy equivalent digital spaces and M is simply connected, then so is N. We show that a simply connected digital 2-manifold is the digital 2-sphere and a simply connected digital 3-manifold is the digital 3-sphere. This property can be considered as a digital form of the Poincaré conjecture for continuous three-manifolds.

## 1. Introduction

A digital approach to geometry and topology plays an important role in analyzing n-dimensional digitized images arising in computer graphics as well as in many areas of science including neuroscience, medical imaging, industrial inspection, geoscience and fluid dynamics. Concepts and results of the digital approach are used to specify and justify some important low-level image processing algorithms, including algorithms for thinning, boundary extraction, object counting, and contour filling.
Usually, a digital object is equipped with a graph structure based on the local adjacency relations of digital points [5]. In papers [6-7], a digital n-surface was defined as a simple undirected graph and basic properties of n-surfaces were studied. Paper [6] analyzes a local structure of the digital space $Z^n$. It is shown that $Z^n$ is an n-surface for all n>0. In paper [7], it is proven that if A and B are n-surfaces and A$\subseteq$B, then A=B.
X. Daragon et al. [3-4] studied partially ordered sets in connection with the notion of n-surfaces. In particular, it was proved that (in the framework of simplicial complexes) any n-surface is an n-pseudomanifold, and that any n-dimensional combinatorial manifold is an n-surface. In paper [14], M. Smyth et al. defined dimension at a vertex of a graph as basic dimension, and the dimension of a graph as the sup over its vertices. They proved that dimension of a strong product G × H is dim ( G ) + dim ( H ) (for non-empty graphs G and H).
An interesting method using cubical images with direct adjacency for determining such topological invariants as genus and the Betti numbers was designed and studied by L. Chen et al. [2].
E. Melin [13] studies the join operator, which combines two digital spaces into a new space. Under the natural assumption of local finiteness, he shows that spaces can be uniquely decomposed as a join of indecomposable spaces. In papers [1, 9], digital covering spaces were classified by using the conjugacy class corresponding to a digital covering space.
A digital n-manifold, which we regard in this paper, is a special case of a digital n-surface studied in [6-8].
In section 3, properties of digital n-spheres, n-manifolds and n-manifolds with boundary are investigated. We show that a digital n-manifold M is a sphere if for any contractible subspace A, the space M-A is also contractible. In section 4, we prove that any normal *(n-1)*-sphere S contained in a normal n-sphere *M* is a separating space in *M*. Sections 5-7 define and study simply connected digital spaces and manifolds. We prove that simply connected digital 2- and 3-manifolds are digital 2- and 3-spheres respectively.

## 2. Preliminaries

A *digital space* G is a simple undirected graph G=(V,W), where V={$v_1,v_2,...v_n,...$} is a finite or countable set of points, and W = {($v_p v_q$),....}⊆V×V is a set of edges provided that ($v_p v_q$)=($v_q v_p$) and ($v_p v_p$)∉W [10].

Such notions as the connectedness, the adjacency, the dimensionality and the distance on a graph G are completely defined by sets V and W. Further on, if we consider a graph together with the natural topology on it, we will use the phrase *'digital space"*. We use the notations $v_p$∈G and ($v_p v_q$)∈G if $v_p$∈V and ($v_p v_q$)∈W respectively if no confusion can result.

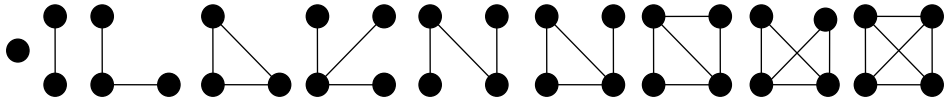

Figure 1. Contractible graphs with the number of points n<5.

Consider a graph H such that V(H) ⊆V(G) and W(H)⊆W(G). Under these conditions, H is called a *subgraph* of G, H⊆G. Note that H is an *induced* subgraph of G if H is obtained by deleting the only points.

The induced subgraph $O(v)$⊆G containing all points adjacent to v (without v) is called *the rim or the neighborhood of point v in G,* the subgraph $U(v)=v\oplus O(v)$ is called *the ball of v*.

For two graphs G=(X,U) and H=(Y,W) with disjoint point sets X and Y, their *join G⊕H* is the graph that contains G, H and edges joining every point in G with every point in H.

Digital spaces (graphs) can be transformed from one into another in a variety of ways. Contractible transformations of graphs (digital spaces) seem to play the same role in this approach as a homotopy in algebraic topology [11-12]. A graph G is called *contractible* (fig. 1), if it can be converted to the trivial graph by sequential deleting simple points. A point v of a graph G is said to be *simple* if its rim O(v) is a contractible graph. An edge (vu) of a graph G is said to be *simple* if the joint rim O(vu)=O(v)∩O(u) is a contractible graph. Deletions and attachments of simple points and edges are called *contractible transformations*. Graphs G and H are called *homotopy equivalent or homotopic* if one of them can be converted to the other one by a sequence of contractible transformations. Homotopy is an equivalence relation among graphs. Contractible transformations retain the Euler characteristic and homology groups of a graph [12].

Properties of graphs that we will need in this paper were studied in [11-12].

### Proposition 2.1
- Let G be a graph and v be a point (v∉G). Then v⊕G is a contractible space.
- Let G be a contractible graph with the cardinality |G|>1. Then it has at least two simple points.
- Let H be an induced contractible subgraph of a contractible graph G. Then G can be transformed into H by sequential deleting simple points.

## 3. Properties of n-sphere and n-manifolds

Since in this section we use only induced subspaces, we use the word subspace for an induced subspace. In paper [6], digital n-surface was defined and investigated. Here we study a special case of n-surfaces: digital n-manifolds.

### Definition 3.1.
A *digital 0-dimensional sphere* is a disconnected graph $S^0$(a,b) with just two points a and b (fig. 2)..

To define n-spheres, n>0, we will use a recursive definition. Suppose that we have defined k-spheres for dimensions $1 \leq k \leq n-1$.

**Definition 3.2.**
- A connected space M is called a *digital n-sphere, n>0,* if for any point $v \in M$, the rim O(v) is an (n-1)-sphere and the space M-v is contractible.
- Let M be a digital n-sphere, n>0, and v be a point belonging to M. The space D=M-v is called *a digital n-disk* (fig. 2, 3).

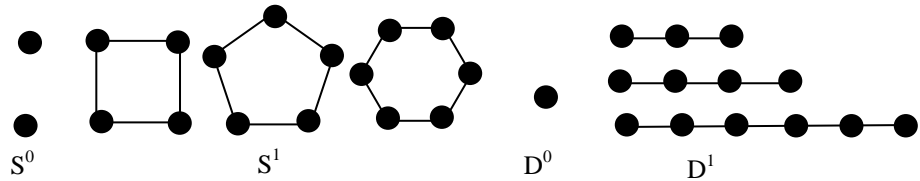

$S^0$  $S^1$  $D^0$  $D^1$

Figure 2. Zero- and one-dimensional spheres $S^0$ and $S^1$ and zero- and one-dimensional disks $D^0$ and $D^1$.

Obviously, D is a contractible space, which can be represented as the union $D=\partial D \cup IntD$ of two non-empty spaces: $\partial D = O(v)$ is a digital (n-1)-sphere, the rim O(x) is a digital (n-1)-sphere if a point x belongs to IntD and the rim O(x) is a digital (n-1)-disk if a point x belongs to $\partial D$ (fig. 2, 3). Subspaces IntD and $\partial D$ are called *the interior* and *the boundary* of D respectively.

It was shown in [7] that the join $S^n_{min} = S^0_1 \oplus S^0_2 \oplus \ldots S^0_{n+1}$ of (n+1) copies of the zero-dimensional sphere $S^0$ is the minimal digital n-sphere (fig. 3). A digital n-sphere M can be converted to the minimal n-sphere $S_{min}$ by contractible transformations. If M is a digital n-sphere, then $S^0(u,v) \oplus M$ is a digital (n+1)-sphere.

**Definition 3.3.**
- A connected space M is called *a digital n-dimensional manifold*, n>1, if the rim O(v) of any point v is a digital (n-1)-dimensional sphere [7].
- A connected space $N = \partial N \cup IntN$ is called a digital n-manifold *with the boundary* $\partial N$ if:
  1. $\partial N$ is a digital (n-1)-manifold.
  2. If a point $x \in IntN$, then the rim O(x) is a digital (n-1)-sphere.
  3. If a point $x \in \partial N$, the rim O(x) is a digital (n-1)-disk.

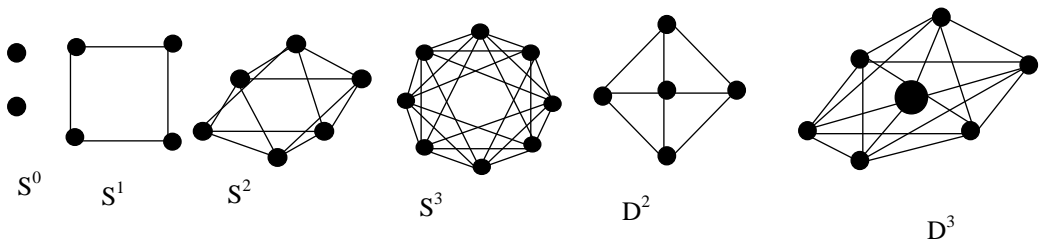

$S^0$  $S^1$  $S^2$  $S^3$  $D^2$  $D^3$

Figure 3. Minimal spheres and disks.

**Proposition 3.1.**
Let M be a digital n-sphere, N be contractible subspace of M and v be a point in M. Then subspaces M-N and M-v are homotopy equivalent to each other.

Proof.
The proof is by induction on the dimension n. For n=1, the proposition is verified directly. Assume that the proposition is valid whenever n<k. Let n=k. Let M be a digital n-sphere and N be a contractible subspace of M.
Since N is contractible, there is a point $x \in N$ such $O(x) \cap N$ is contractible (proposition 2.1). Since O(x) in M is a digital (n-1)-sphere, then by the induction hypothesis, $O(x)-N=O(x) \cap (M-N)$ is also contractible. Therefore, C=N-x is a contractible space and E=(M-N)∪x  is homotopic to M-N. Acting in the same way,  we finally convert the space N to a point v and the space M-N to the  space  M-v. Since M-v is contractible then M-N is a  contractible space. This completes the proof. □

**Theorem 3.1.**
>Let M  be a digital n-manifold, N be a contractible  subspace of  M and v be a point in M. Then subspaces M-N and M-v are  homotopy equivalent to each other.

The proof of this statement is similar to the proof of  proposition 3.1 and is omitted here.
The following corollary is an obvious consequence of propositions 3.1 and theorem 3.1.

**Corollary 3.1.**
>Let M  be a digital n-manifold  and N be a contractible  subspace of  M. M is a digital n-sphere if and only if  the space M-N is contractible.

**Proposition 3.2.**
>Let D=∂D∪IntD be a digital n-disk with the boundary  ∂D and  X be contractible subspace of ∂D. Then any point v belonging to the space Y=∂D-X  is simple in F=D-X=(∂D-X)∪IntD  and the space  F  is homotopy equivalent to D.  In particular IntD is homotopy equivalent to D.

Proof.
Notice first that according to definition 3.2, if v∈∂D then O(v)=∂E∪IntE is an (n-1)-disk with the boundary ∂E=O(v)∩∂D and the interior IntE=O(v)∩IntD.
The proof is by induction on the dimension n. For n=1, the proposition is plainly true.
Assume that the proposition is valid whenever n<k. Let n=k. Let v∈∂D-X. Then O(v) in F is the space O(v)∩F=O(v)∩((∂D-X)∪IntD)=(O(v)∩(∂D-X))∪(O(v)∩IntD)=(∂E-X))∪IntE. By the inductive assumption, O(v)∩F is homotopy equivalent to O(v)=∂E∪IntE. Since O(v) is a contractible space then O(v)∩F is a contractible space. Hence, v is a simple point in F and can be deleted from F. For the same reason by deleting all points belonging to X (all of these points are simple), we obtain F, which is homotopy equivalent to D. Similarly, IntD is homotopy equivalent to D.   This completes the proof. □

**Theorem 3.2**.
>Let N=∂N∪IntN  be  a digital n-manifold *with the boundary*  ∂N and  X be contractible subspace of  ∂N. Then any point v belonging to the space Y=∂N-X is simple in F and  the space F=N-X=(∂N-X)∪IntN  is homotopy equivalent to N.  In particular IntN is homotopy equivalent to N.

The proof of this theorem is similar to the proof of the previous  proposition  and so  is omitted.

**Proposition 3.3.**
>Let M be a contractible space  and a digital 1-sphere S be a subspace of  M. Then by sequential deleting simple point and edges,  M is converted to  a digital 2-disk D=∂D∪IntD  with the boundary   S=∂D.

Proof.
Sequentially delete from M simple points and edges, which do not belong to S. In the obtained space D, if a point x∉S and an edge  (xy)∉S, then   x and (xy) are not simple.  Suppose that a point x∈S. Then S-x is a contractible space. Therefore, D can be converted to S-x by sequential deleting simple points (proposition 2.1).   This means that x is simple in D,  i.e. O(x) is a contractible space. O(x) contains nonadjacent points y and z belonging to S and adjacent to x. Therefore, there is a path L=L(y,z) (digital 1-disk)  with endpoints y and z, L⊆O(x). Assume that L≠O(x). Since O(x) can be converted to L be sequential deleting simple points (proposition 2.1), then there is a point v∈O(x), x∉L, which is simple in O(x), i.e. O(vx) is a contractible space. This means that the edge (xv) is simple. This is a contradiction. Hence, the assumption is false and O(x)=L.
Glue a point p to D-x , O(p)=S-x. The obtained space $D_1$=(D-x)∪p is contractible, contains the digital

1-sphere $S_1=L\cup p$ and does not contain simple points and edges except for lying in $S_1$. For the same reason as above, for any point in $S_1$, say, $u\in L$, $u\neq y,z$, $O(u)$ in $D_1$, i.e., $O(u)\cap D_1$ is a digital 1-disk $L_1$ in $D_1$. Hence, $O(u)$ in $D$ is a digital 1-sphere $L_1\cup x$ by construction.
By repeating the same procedure, we finally show that the rim of any point $v\in D-S$ is a 1-sphere, i.e., D is a digital 2-disk. □
The following assertion was proven in [7].

**Proposition 3.4.**
  Let M and N be a digital n-manifolds. If N is a subspace of M, then M=N.

## 4. Separation of spaces

Since in this section we use only induced subspaces, we use the word subspace for an induced subspace.

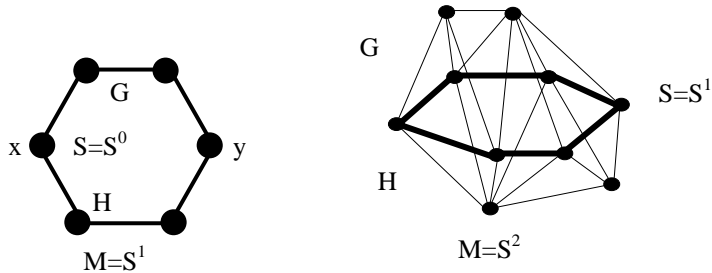

Figure 4. A 0-sphere $S^0=\{x,y\}$ is a separating space in a 1-sphere M. A 1-sphere $S^1$ is a separating space in a 2-sphere $M=S^2$.

**Definition 4.1.**
  Let A and B be subspaces of a connected space M. A and B are called *separated* if any point in A is non-adjacent to any point in B.
  If a space M is represented as the union $A\cup C\cup B$, where spaces A and B are separated, we will say that the union $M=A\cup C\cup B$ *is a separation* of M by the space C and C is a *partition for M*.
It is easy to prove the following property.

**Proposition 4.1.**
  Let $M=A\cup C\cup B$ be the separation of a space M by a contractible subspace C. M is contractible if and only if subspaces $A\cup C$ and $C\cup B$ are both contractible.

**Proposition 4.2.**
   Let M be a digital n-sphere and S be a digital (n-1)-sphere in M, $S\subseteq M$. Then $M=G\cup S\cup H$ is the separation of M by S and $G\cup S$ and $S\cup H$ are n-disks (fig. 4).
Proof.
The proof is by induction on the dimension n. For n=1, the proposition is plainly true (fig. 4). Assume that the proposition is valid whenever n<k. Let n=k. Since $S\subset M$, there is a point v, belonging to M and not belonging to S. E=M-v is a digital n-disk and $S\subseteq E$. Sequentially delete from E simple points which do not belong to S. We obtain a contractible space D, where any point belonging to D-S, is not simple. Let a point $x\in S$. Since S-x is a contractible subspace of D, then x is a simple point in D according to proposition 2.1. By construction, $C=O(x)\cap S$ is a digital (n-2)-sphere. By the induction hypotheses, $O(x)=A\cup C\cup B$ is the separation of $O(x)$ by C and $O(x)\cap D=C\cup B$ is a digital (n-1)-disk. Glue to D a point y, where $O(y)=S-x$, and delete x. The obtained space $D_1=(D-x)\cup y$ is contractible and $C\cup B\cup y$ is a digital (n-1)-sphere in $D_1$. For the same reason as above, for any point $x_1\in B$, $O(x_1)\cap D_1= C_1\cup B_1$ is a digital (n-1)-disk. Therefore, $O(x_1)\cap D= x\cup C_1\cup B_1$ is a digital (n-1)-sphere contained in D. According to proposition 3.4, $O(x_1)\cap D=O(x_1)\cap M=O(x_1)$. Thus, $x_1$ is not adjacent to any point $q\in M-D$. Using the same procedure we show that for any point $p\in D-S$, $O(p)\subseteq D$. This means that D is an n-disk and $M=(M-D)\cup S\cup(D-S)$ is a separation of M by S. The proof is complete. □

**Corollary 4.1.**
Let $D=\partial D\cup IntD$ and $E=\partial E\cup IntE$ be digital n-disks. If $\partial D$ and $\partial E$ are isomorphic, f: $\partial D\rightarrow \partial E$, then the space D#E obtained by identifying points belonging to $\partial D$ with corresponding points belonging to $\partial E$ is a digital n-sphere.

## 5. Properties of simply connected spaces and manifolds

**Definition 5.1.**
A connected space C is called a *closed curve* if for any point v in C, the rim O(v) is a disconnected graph with just two points (fig. 5).
In other word, a closed curve is a digital 1-sphere.

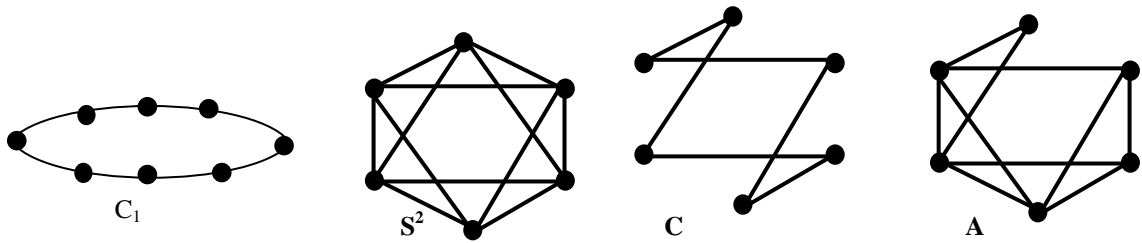

Figure 5. $C_1$ and C are closed curves. $S^2$ is a 2-sphere. A is a contractible space. C $S^2$, A $S^2$, C A.

**Definition 5.2.**
A connected space M is called *simply connected* if whenever C is a closed curve obtained from M by deleting points and edges, then there is a contractible space A obtained from M by deleting points and edges and such that C is obtained from A by deleting points and edges, i.e. $C\subseteq M$, $A\subseteq M$, $C\subseteq A$.

A digital 2-sphere $S^2$ is depicted in fig. 5. A closed curve C is obtained from $S^2$ by deleting edges. A contractible space A is obtained from $S^2$ by deleting two edges. Evidently, C is obtained from A by deleting edges, $C\subseteq A$. It can be checked directly that for any closed curve $C\subseteq S^2$ there is a contractible space $A\subseteq G$ such that $C\subseteq A$. Therefore, $S^2$ is a simply connected space.

**Proposition 5.1.**
Let M be a space and v be a simple point in M. M is simply connected if and only if M-v is simply connected.
Proof.
(A) Let M-v be simply connected. Suppose that C is a closed curve in M.
(A1) Assume that $v\notin C$. Then C is a closed curve in M-v and there is a contractible space A, obtained from M-v by deleting points and edges and $C\subseteq A$. Evidently, $A\subseteq_g M$.
(A2) Let $v\in C$. With no loss of generality, assume that the induced space containing point belonging to $O(v)\cap C$ is a disconnected graph with just two points x and y, $O(v)\cap C=\{x,y\}$ (fig. 6). Since O(v) is contractible, then there is a path L(x,y), $L(x,y)\subseteq O(v)$. Consider the induced subspace $(C-v)\cup L(x,y)$ and delete from $(C-v)\cup L(x,y)$ all edges connecting points p and q, where $p\in C-v-x-y$, $q\in L(x,y)-x-y$. The obtained space $C_1$ is a closed curve by construction. Therefore, there is a contractible space A (fig. 6), obtained from M-v by deleting points and edges and containing $C_1$. Then the space $B=v\cup A$, where $O(v)\cap B=L(x,y)$ is a contractible space containing C (fig. 6). Hence, M is a simply connected space.
(B) Let M be a simply connected space. Suppose that a closed curve C is obtained from M-v by

deleting points and edges. Then there is a contractible space A obtained from M by deleting points and edges and A contains C, C⊆A.

(B1) If the point v does not belong to A, then A⊆M-v.

(B2) Assume that v∈A. Consider the induced union B=A∪O(v) (fig. 7). Delete from B all edges (xy), where y∈A-O(v), x∈O(v)-A. The obtained space $B_1$ contains the point v, which is simple in $B_1$. Therefore, $B_2=B_1-v$ is homotopy equivalent to $B_1$. On the other hand, any point x belonging to O(v)-A is also simple in $B_1$, since $O(x) \cap B_1 = v \otimes (O(xv))$. Therefore, any such point can be deleted from $B_1$. The obtained space is the space A, which is homotopy equivalent to $B_1$ and, therefore, $B_2$. Hence, $B_2$ is a contractible space. Since $B_2$ does not contain the point v, then M-v is a simply connected space. The proof is complete. □

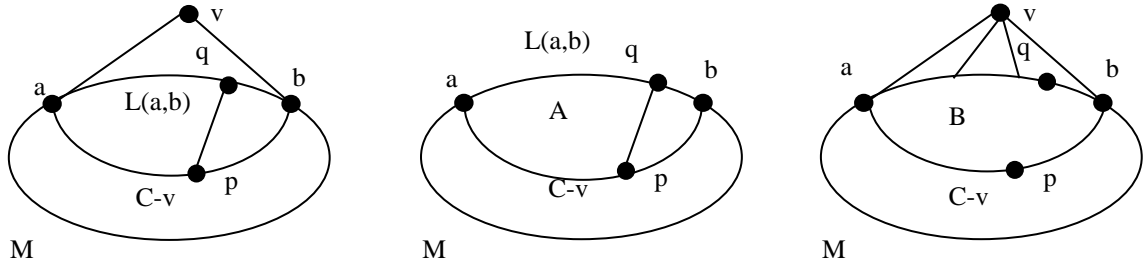

Figure 6. M is a simply connected space, C is a closed curve, v is a simple point in M. B is a contractible space in M-v. $B_1$ is a contractible space containing C.

**Proposition 5.2.**
   Let M be a space and (vu) be a simple edge in M. M is simply connected if and only if M-(vu) is simply connected.

The proof of this proposition is similar to the proof of proposition 5.1 and is omitted here.
The following theorem is a direct consequence of propositions 5.1 and 5.2.

**Theorem 5.1.**
   Let spaces M and N be homotopy equivalent. M is simply connected if and only if N is simply connected.

The proof of the following theorem is similar to the proof of proposition 5.1.

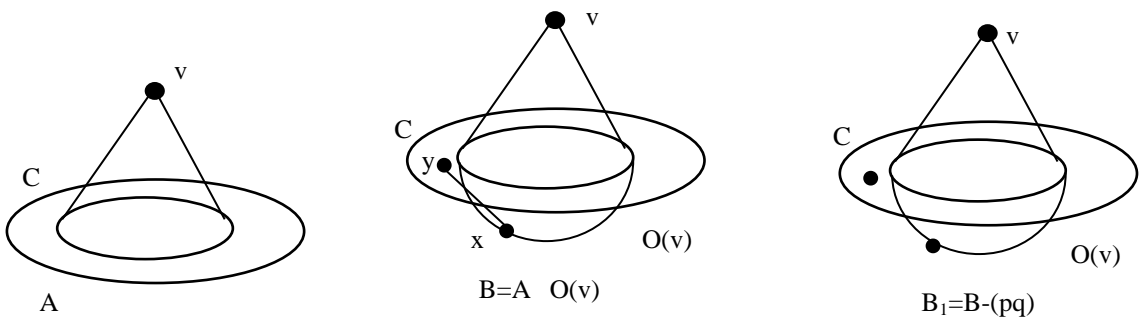

Figure 7. A is a digital 2-disk. B O(v) contains the edge (xy). B O(v)-(xy) is a contractible space.

**Theorem 5.2.**
   Let M be a digital n-manifold, n>1, and v be a point in M. M is simply connected if and only if M-v is simply connected.
Proof.

(A) Let the space M-v is simply connected. Suppose that C is a closed curve obtained from M by deleting points and edges.
(A1) Assume that v∉C. Then there is a contractible subspace A obtained from M-v by deleting points and edges such C⊆A. Evidently, A⊆M.
(A2) Assume that v∈C. With no loss of generality, assume that the induced space containing point belonging to O(v)∩C is a disconnected graph with just two points x and y, O(v)∩C={x,y} (fig. 6). Since the induced subspace O(v) is an (n-1)-sphere, then there is a path L(x,y), L(x,y)⊆O(v). Delete from the space (C-v)∪L(x,y) all edges connecting points p and q, where p∈C-v-x-y, q∈L(x,y)-x-y. The obtained space $C_1$ is a closed curve by construction. Therefore, there is a contractible space A containing $C_1$. (fig. 6) and obtained from M-v by deleting points and edges. Then the space B=v∪A, where O(v)∩B=L(x,y) is a contractible space containing C (fig. 6). Hence, M is a simply connected space.
(B) Let M be a simply connected space. Suppose that a closed curve C is obtained from M-v by deleting points and edges and a contractible space A obtained from M by deleting points and edges contains C, C⊆A.
(B1) If the point v does not belong to A, then A⊆M-v.
(B2) Assume that v∈A. Consider the union B=A∪O(v) (fig. 7). Delete from B all edges (xy), where y∈A-O(v), x∈O(v)-A. The obtained space $B_1$ contains the point v, which is simple in $B_1$. Therefore, $B_2=B_1-v$ is homotopy equivalent to $B_1$. On the other hand, any point x belonging to O(v)-A is also simple in $B_1$, since $O(x)\cap B_1=v\otimes(O(xv)$. Therefore, any such point can be deleted from $B_1$. The obtained space is the space A, which is homotopy equivalent to $B_1$ and, therefore, $B_2$. Hence, $B_2$ is a contractible space. Since $B_2$ does not contain the point v, then M-v is a simply connected space. The proof is complete. □

**Corollary 5.1.**
Let M be a digital n-manifold, n>1, and N be a contractible induced subspace of M. Them M is simply connected if and only if the space M-N is simply connected.
This statement follows directly from theorem 5.1, theorem 5.2 and theorem 3.1.

**6. Properties of simply connected 2-manifolds**

For simplicity, all closed curves considered in this section are induced digital 1-spheres.

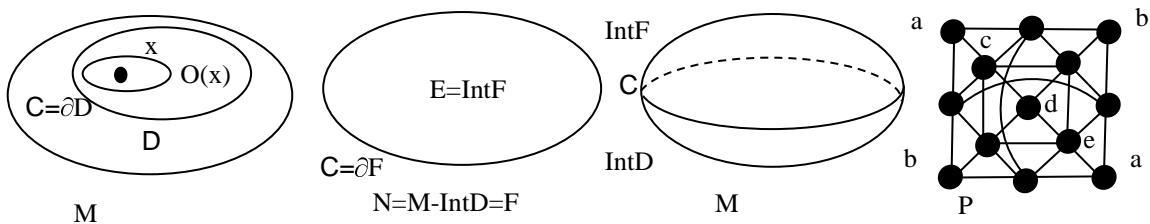

Figure 8. M is a simply connected 2-manifold, C is a closed curve in M, D is a 2-disk with the boundary C. D is a subspace of M. M=E C IntD is a separation of M by C. N=M-IntD= E C=F is a digital 2-disk with the boundary C=∂F; M=IntF C IntD. P is a digital projective plane. For a closed curve C={a,c,d,e}, there is no contractible subspace of P containing C.

**Proposition 6.1.**
Let M be a simply connected digital 2-manifold. Then M is a digital 2-sphere.
Proof.
Let C be a 1-sphere contained in M. Then there is a contractible space D containing C and obtained from M by deleting edges and points. With no loss of generality, consider D=∂D∪IntD as a digital 2-disk with the boundary C=∂D according to proposition 3.3.
Let a point x belong to IntD (fig. 8). Then the rim O(x) of x in D is a digital 1-sphere $S_1$=O(x) ∩D. On the other hand, the O(x) in M is also a digital 1-sphere S=O(x). Since $S_1$⊆S, then $S_1$=S according to proposition 3.4. . Therefore, O(x)⊆D. Hence, C is a separating space in M, D is an induced subspace of M and M=E∪C∪IntD is a separation of M by C.
Since IntD is a contractible space, then N=M-IntD=E∪C (fig.8) is a simply connected space according corollary 5.1. Therefore, there is a contractible space F containing C and obtained from N by deleting edges and points. As above, consider F=∂F∪IntF as a digital 2-disk with the boundary C=∂F. Since O(x)∩F=O(x) ∩N for any point x, x∈E, then N=E∪C=F, i.e. a contractible space. Therefore, M=IntE∪C∪IntD is a digital 2-sphere according to corollary 3.1. □

The digital projective plane P depicted in fig. 8 is not simply connected. It is checked directly that for a closed curve C={a,c,d,e} there is no contractible space A, A⊆P such that C⊆A.

**Properties of simply connected 3-manifolds**

Note that all closed curves considered in this section are induced digital 1-spheres.

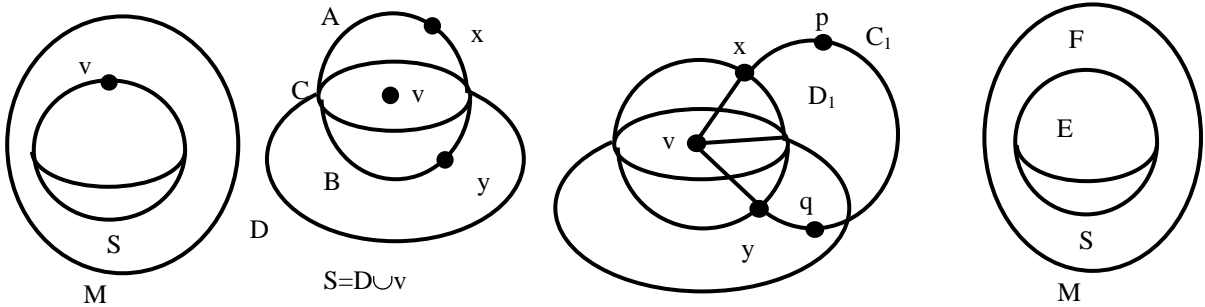

Figure 9. M is a simply connected 3-manifold, S is a 2-sphere, S  M. D=S-v, O(v)=A∪C∪B, a closed curve $C_1$ intersects D. M=E∪S∪F is a separation of M by S.

**Theorem 7.1.**
Let M be a simply connected digital 3-manifold and S be a digital induced 2-sphere contained in M. Then S is a separating space in M=E∪S∪F and spaces E∪S and S∪F are simply connected.

Proof.
Pick a point v belonging to S. Then C=O(v)∩S is a 1-sphere, O(v)=A∪C∪B is a separation of O(v) by C and A and B are contractible spaces (fig. 9) according to proposition 4.2 and corollary 3.1. Since the space D=S-v is a digital 2-disk, i.e. a contractible space (fig. 9), then M-D is a simply connected space according to corollary 5.1. Pick points x∈ A, y∈ B and consider a closed curve $C_1$ containing points x, v, y. Assume that $C_1$ does not intersect D. Then there is a contractible space $D_1$ in M-D containing $C_1$ and obtained from M-D by deleting point and edges. With no loss of generality, consider $D_1$=∂$D_1$∪Int$D_1$ as a digital 2-disk with the boundary $C_1$=∂$D_1$ according to proposition 3.3. Since v∈ $D_1$ and C⊆D, then D∩$D_1$≠∅ (fig. 9). Therefore, the assumption is false and $C_1$ must intersect D, D∩$C_1$≠∅. Pick points p and q belonging to $C_1$ and paths L(p,x)⊆$C_1$, L(q,y)⊆$C_1$, L(p,q)⊆$C_1$ such that D∩L(p,x)=∅, D∩L(q,y)=∅, v∉L(p,q). Then L(p,q) must intersect D.
Let E be a set of points such that if p∈ E then there is a path L(p,x) not intersecting D and F be a set of points such that if q∈F then there is a path L(q,y) not intersecting D. It follows from the above that any path L(p,q) must intersect S (for example in the point v). Therefore, S is a separating space in M=E∪S∪F (fig. 9). Hence, spaces E∪S and S∪F are both digital 3-manifolds with the boundary S by definition 3.3.
M-D=E∪v∪F is a simply connected space according to corollary 5.1. Then spaces E∪v and v∪F are both simply connected by construction. Since v is a simple point in E∪v and v∪F, then E and F are both simply connected by theorem 5.1. Therefore, spaces E∪S and S∪F are both simply connected according to theorem 3.2 and theorem 5.1. □

An R-transformation is the replacement of an edge (xy) with a point z in a such a manner that the rim $O(z)=S^0(x,y)\otimes O(xy)$. An R-transformation is a sequence of two contractible transformations. The following assertion was proven in [8].

**Proposition 7.1.**
Let N=RM is a space obtained from M by R-transformations. Spaces M and N are homotopy equivalent. M is an n-manifold if and only if N is an n-manifold.

**Theorem 7.2.**
If M is a digital simply connected 3-manifold, then M is the digital 3-sphere.
Proof.
The proof is by induction on the number of points n=|M| in M. For small n=8,,..., the theorem is verified directly (fig. 3). Assume now that the theorem is valid whenever n<k. Let n=k.
(A1) Let $v\in M$, $C\subseteq O(v)$ be a digital 1-sphere, $O(v)=P\cup C\cup Q$ be the separation of O(v) by C. Since the space $E=v\oplus(P\cup Q)$ is contractible, than M-E is a simply connected space according corollary 5.1.
Then there is a contractible space D in M-E containing C and obtained from M-E by deleting point and edges. With no loss of generality, consider $D=\partial D\cup IntD$ as a digital 2-disk with the boundary $C=\partial D$ (fig. 10) according to proposition 3.3.

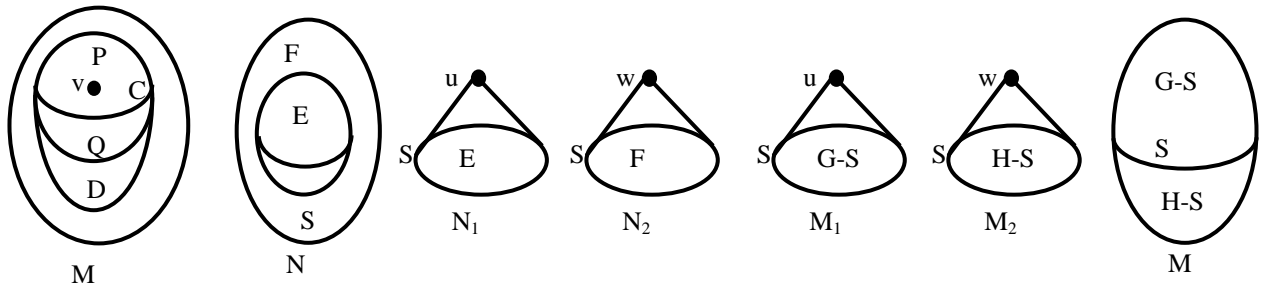

Figure 10. M is a simply connected 3-manifold. S  N is a 2-sphere separating N. $N_1$ and $N_2$ are digital 3-spheres. $M_1=R^{-1}N_1$ and $M_2=R^{-1}N_2$ are digital 3-spheres. M= $(G-S)\cup S\cup(H-S)$ is a digital 3-sphere.

(A2) Let D be obtained from M-E by deleting points $p_1,...p_k$ and edges $e_1,...e_m$.
Replace any $e_i=(x_iy_i)$ with a point $z_i$. According to proposition 7.1, the obtained space N=RM is an n-manifold homotopy equivalent to M. Therefore, N is simply connected by theorem 5.1. Evidently, D is obtained from N by deleting points $p_1,...p_k$ and $z_1,...z_m$ and $S=D\cup v$ is a digital 1-sphere contained in N. Then S is a separating space in $N=E\cup S\cup F$, spaces $E\cup S$ and $S\cup F$ are simply connected digital 3-manifolds with the boundary S (fig. 10) according to theorem 7.1 and $P\subseteq E$, $Q\subseteq F$. By construction, points $p_i$ and $z_i$ belong to either E or F.
(A3) With no loss of generality, assume |E|>1, |F|>1 and consider spaces $N_1=E\cup S\cup u$, $N_2=w\cup S\cup F$, obtained by gluing points u and w to $E\cup S$, $S\cup F$ respectively, where O(u)=S, O(w)=S (fig. 10).
$N_1$ and $N_2$ are simply connected 3-manifolds by theorem 5.2. Apply to $N_1$ and $N_2$ the inverse R-transformations, i.e. replace any point $z_i$ with the edge $(x_iy_i)$. By proposition 7.1, the obtained spaces $M_1=R^{-1}(N_1)=R^{-1}(E\cup S)\cup u$ and $M_2=R^{-1}(N_2)=w\cup R^{-1}(S\cup F)$ are digital 3-manifolds homotopy equivalent to $N_1$ and $N_2$ (fig. 10). $M_1$ and $M_2$ are simply connected 3-manifolds by theorem 5.2. Since the number of points $|M_1|<n$ and $|M_2|<n$, then $M_1$ and $M_2$ are digital 3-spheres by the induction hypothesis. Therefore, $G=M_1-u$ and $H=M_2-w$ are digital 3-disks with the boundary S according to definition 3.2. Since $M=(G-S)\cup S\cup(H-S)$ (fig. 10), then M is a digital 3-sphere according to corollary 4.1. □

This theorem is a digital counterpart of the Poincaré conjecture for continuous three-manifolds.
A link between a continuous and a digital 3-manifold can be established by using the intersection graph of a cover of the 3-manifold studied in [7].